\documentclass{PoS}

\title{New results from the DANSS experiment}

\ShortTitle{New results from the DANSS experiment}

\author{\speaker{Mikhail Danilov}\\
        Lebedev Physical Institute of the Russian Academy of Sciences,\\
        53 Leninskiy Prospekt, Moscow, Russia\\
        E-mail: \email{danilov@lebedev.ru}}

\author{Nataliya Skrobova\\
        Lebedev Physical Institute of the Russian Academy of Sciences,\\
        53 Leninskiy Prospekt, Moscow, Russia\\
        Alikhanov Institute for Theoretical and Experimental Physics NRC "Kurchatov Institute",\\ B. Cheremushkinskaya 25, Moscow, 117218, Russia\\On behalf of the DANSS Collaboration
        }

\abstract{The DANSS experiment collected 5.5 million Inverse Beta Decay (IBD) events during 5 years of operation. Data were collected at 3 distances (10.9~m, 11.9~m, and 12.9~m) from the  center of the core of the industrial $3.1~GW$ reactor. The IBD event rate exceeds 5000/day at the smallest distance. The detector position was changed usually 3 times a week. Therefore in the analysis that uses information only about relative IBD counting rates and changes in positron energy spectra shapes many systematic uncertainties were canceled out.  No statistically significant evidence for sterile neutrinos is found. 
The  significance of the best-fit point in the 4$\nu$ case is only 1.3$\sigma$. 
The excluded area covers a very interesting range of the sterile neutrino parameters up to $sin^22\theta_{ee} < 0.008$ in the most sensitive region.
The IBD rate dependence on the fission fraction of $^{239}$Pu was measured. It agrees with predictions of the Huber-Mueller model. The reactor power was measured during more than 4 years with $\approx2\%$ accuracy in 2 days  using the rate of neutrino events normalized to the thermal power at the initial period.
}

\FullConference{%
  *** The European Physical Society Conference on High Energy Physics (EPS-HEP2021), ***\\
  *** 26-30 July 2021 ***\\
  *** Online conference, jointly organized by Universität Hamburg and the research center DESY ***
}

\begin{document}

\def\mydm{\Delta m^2_{41}}
\def\mysin{\sin^2 2\theta_{ee}}
\def\oscillationspars{$\mydm$, $\mysin$}
\def\antiparticle{\tilde}
\def\antinu{$\tilde{\nu_e}$}
\def\clsmethod{CL$_s$}
\def\cl{\mathrm{CL}}

\section{Introduction}

Measurements of the Z boson width fix the  number of light active neutrinos to 3. However, additional sterile neutrinos are not excluded and    there are several experimental indications of their existence.
The deficit of $\nu_e$ in the calibration of the SAGE and GALEX experimets with radioactive sources~\cite{SAGE, GALEX} (``Galium Anomaly''(GA)) and the deficit in reactor $\antiparticle\nu_e$ fluxes~\cite{Huber,Mueller} (``Reactor Antineutrino Anomaly''(RAA)) can be explained by active-sterile neutrino oscillations with the  best-fit values for sterile neutrino parameters of $\Delta m^2_{41} = 2.4$ eV$^2$ and $\sin^2 2\theta_{ee}$ = 0.14 ~\cite{Ga,Mention2011}. 

New measurements~\cite{Kopeikin:2021rnb,Kopeikin:2021ugh} of beta spectra of fission products of $^{235}$U and $^{239}$Pu give 5.4\%  smaller ratio than the ILL results used for predictions of the $\antiparticle\nu_e$ fluxes from reactors~\cite{Huber,Mueller}. This can explain RAA. However, all modern searches for sterile neutrinos do not rely on the absolute flux measurements.

Very recently the BEST experiment confirmed GA~\cite{Barinov:2021asz}. The deficit of $\nu_e$ events becomes even larger ($20\pm 5\%$) and has a larger significance of about 5$\sigma$ (this is our naive estimate not given in the paper). For $\Delta m^2_{41} < 5$eV$^2$ very large values of $\sin^2 2\theta_{ee}\approx 0.4$ preferred by BEST are excluded by DANSS~\cite{Alekseev:2018efk,Danilov:2020ucs} and NEOS~\cite{Ko:2016owz}. For $\Delta m^2_{41} > 5$eV$^2$ the BEST results are in tension with with limits obtained by Daya Bay, Bugey-3 and RENO (see for example~\cite{MINOS:2020iqj}) using predictions for the absolute $\antiparticle\nu_e$ flux from reactors.

The MiniBooNE collaboration obtained  a 4.8$\sigma$ evidence for $\nu_e$($\antiparticle\nu_e$) appearance in the 
$\nu_{\mu}$($\antiparticle\nu_{\mu}$)
beams~\cite{MiniBooNE2}. A combination of these results with  
earlier 
LSND results has a 6$\sigma$ significance. Appearance of electron neutrinos in muon neutrino beams can be explained by oscillation of muon neutrinos to sterile neutrinos and then to electron neutrinos. 

The Neutrino-4 experiment claimed an observation of $\antiparticle\nu_e$ oscillations to sterile neutrinos  
although the significance of the result was only 2.8$\sigma$\cite{Serebrov:2018vdw,Serebrov:2020rhy} and there were concerns about the validity of their analysis~\cite{Danilov:2018dme,Danilov:2020rax,Almazan:2020drb,Giunti:2021iti}. Neutrino-4 replied to these critical comments\cite{Serebrov:2020wny,Serebrov:2020yvp} and later agreed with two of them\cite{Serebrov:2020kmd}. This resulted in the reduction of the significance of the result by about 0.5$\sigma$. The best-fit point for the increased data sample and improved analysis is  $\Delta m^2_{41} = 7.3\pm 1.17$ eV$^2$ and $\sin^2 2\theta_{ee} = 0.36\pm0.12_{\rm stat}$\cite{Serebrov:2020kmd}. The statistical significance of the result is 2.7$\sigma$. The Neutrino-4 results are in tension with PROSPECT measurements~\cite{Andriamirado:2020erz} as well as with the measurements of the absolute $\antiparticle\nu_e$ flux from reactors~\cite{MINOS:2020iqj}. On the other hand the Neutrino-4 results are in a perfect agreement with the recent BEST result~\cite{Barinov:2021asz}.

The survival probability of reactor $\antiparticle\nu_e$ at very short distances where known oscillations can be neglected in the 4$\nu$ mixing scenario (3 active and 1 sterile neutrino) is given by the formula:\\
$1-\mysin \sin^2({1.27\mydm [\mathrm{eV}^2] L[\mathrm m]}/{E_\nu [\mathrm{MeV}]}),$
where $\mysin$ is the mixing parameter, $\mydm = m_4^2 - m_1^2$ is the difference in the squared masses of neutrino
mass states, $L$ is the distance between production and detection points and $E_{\nu}$ is the $\antiparticle\nu_e$ energy.
 The Inverse Beta Decay (IBD) reaction 
$\antiparticle{\nu}_e + p \rightarrow e^+ + n$ 
is used to detect $\antiparticle\nu_e$. In this reaction $E_{\nu}\approx E_{e^+} + 1.8~MeV$.
\section{Energy calibration, backgrounds and positron spectra.}
The DANSS detector~\cite{DANSS} is located on a movable platform under 
an industrial 3.1~GW$_{th}$ reactor of the Kalininskaya NPP in Russia.
 It consists of 2500 $100\times4\times1~cm^3$ scintillator strips with Gd-loaded surface coating. Strips in neighbor layers are orthogonal which allows a quasi-3D reconstruction of events.  Each strip is readout with 
3 wavelength-shifting fibers placed in grooves along the strip. The central
fiber is read out by a SiPM and the two side fibers from 50
strips are readout by a  PMT. The scintillator strips are
surrounded by a multi-layer passive and active shielding. 
 The kinetic energy  of reconstructed positron is used in the analysis without adding energies of annihilation gammas that suffer from a nonlinear energy response in all experiments. 
A positron kinetic energy is 1.02~MeV smaller than the prompt energy used in other experiments. 

The SiPM's gains and cross-talks are calibrated every 25-30 minutes using noise signals. The energy responses of all 2500 scintillator strips are calibrated every 2 days using cosmic muons. In the present analysis we use a median of the Landau distribution which is more stable than the most probable value used in previous analyses. 
The detector energy scale is anchored by the  $^{12}$B beta spectrum which like the positron tracks does not suffer from a non-linearity in the energy response typical for radioactive sources with many soft gammas. 
The energy scale determined using  $^{22}$Na, $^{60}$Co, and $^{248}$Cm radioactive sources agrees within 1\% with the $^{12}$B scale.  However, we still keep a conservative estimate of 2\% for the energy scale uncertainty in the analysis.

 Observed energy resolution for different calibration sources is slightly worse than MC predictions (33\% instead of 31\% at 1 MeV). Therefore additional smearing is added to the MC predictions ($\sigma_{additional}/E = 12\%/\sqrt{E} \oplus 4\%$). With this correction MC describes well all calibration sources.

The accidental coincidence background is the largest one in the experiment but it is calculated without any model dependence using  time intervals for neutron signals shifted back in time with respect to the $e^+$ signal. 
Several cuts are used to suppress this background~\cite{DANSSdata,Svirida:2020zpk}.
They reduce the accidental background to 15.4\% of the IBD signal at the top detector position for $E_{e^+}$ in (1.5-6)~MeV range used for oscillation analysis. This position and energy range are used as a reference for other background estimates below.
The background from neighbor reactors has a well known shape and constitutes 0.6\%.
Other correlated backgrounds were estimated using reactor-off periods. They constitute less than 1.7\%.
This corresponds to the
signal/background ratio of more than 50. 

\begin{figure}[h]
\begin{center}
\includegraphics[width=0.99\textwidth]{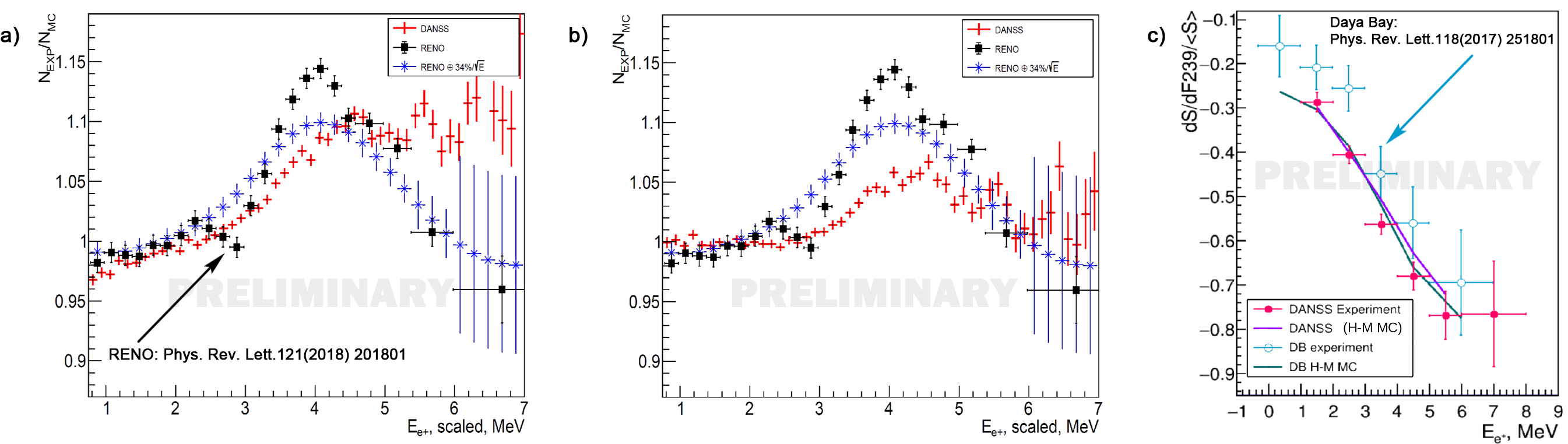} 
\end{center}
\caption{\label{fig:bump} a) Experiment to MC ratio for $e^+$ spectrum; b) the same ratio with energy scaled by 0.995 and shifted by -25~keV; c) relative IBD rate dependence on the $^{239}$Pu fission fraction for different $e^+$ energies; in case of DANSS errors are statistical only.}
\end{figure}

A ratio of the measured $e^+$ spectrum and MC predictions  is shown in Fig.~\ref{fig:bump}a.
The MC simulations assume the Huber-Mueller (H-M) $e^+$ spectrum~\cite{Huber,Mueller}.
RENO results
~\cite{Bak:2018ydk} 
are shown
for comparison shifted by -1.02~MeV since DANSS uses the $e^+$ kinetic energy without annihilation gammas. The RENO spectrum smeared by the DANSS
energy resolution is also shown. The best agreement between our measurements and MC predictions
in the range $1.5 - 3.0$~MeV is obtained if $e^+$ energy is shifted by -25~keV and scaled by 0.995 (Fig.~\ref{fig:bump}b).
Such a change in the energy scale is certainly within the systematic uncertainties. However, we do not know reasons for these particular values and
therefore we can not prove the existence  of the bump.  
Fortunately the search for sterile neutrinos does not depend on the exact shape of the $e^+$ spectrum since only ratios of $e^+$ spectra at the different distances from the reactor core are considered.  

Fig.~\ref{fig:bump}c shows the relative IBD rate dependence on the $^{239}$Pu fission fraction for different $e^+$ energies. The results are in agreement with the H-M model MC predictions and slightly larger in absolute values than the Daya Bay results
~\cite{An:2017osx}
. 
Fig.~\ref{fig:exclusion}a 
shows a comparison of the reactor power and IBD event rate corrected for the changes in the efficiency and fuel composition according to the H-M model. The rate at different detector positions was scaled as $1/L^2$. It was normalized to the reactor power in 1 month period in 2016. After that the $\antiparticle\nu$ counting rate in about 2 day periods coincides with the reactor power within $\sigma=1.9\%$ for more than 4 years (see Fig.~\ref{fig:exclusion}a). Thus the reactor power was measured with $\antiparticle\nu$ counting rate with the accuracy of $\approx2\%$ in 2 days.

\section{Search for sterile neutrinos.}

\begin{figure}[h]
\centering
\includegraphics[width=0.99\textwidth]{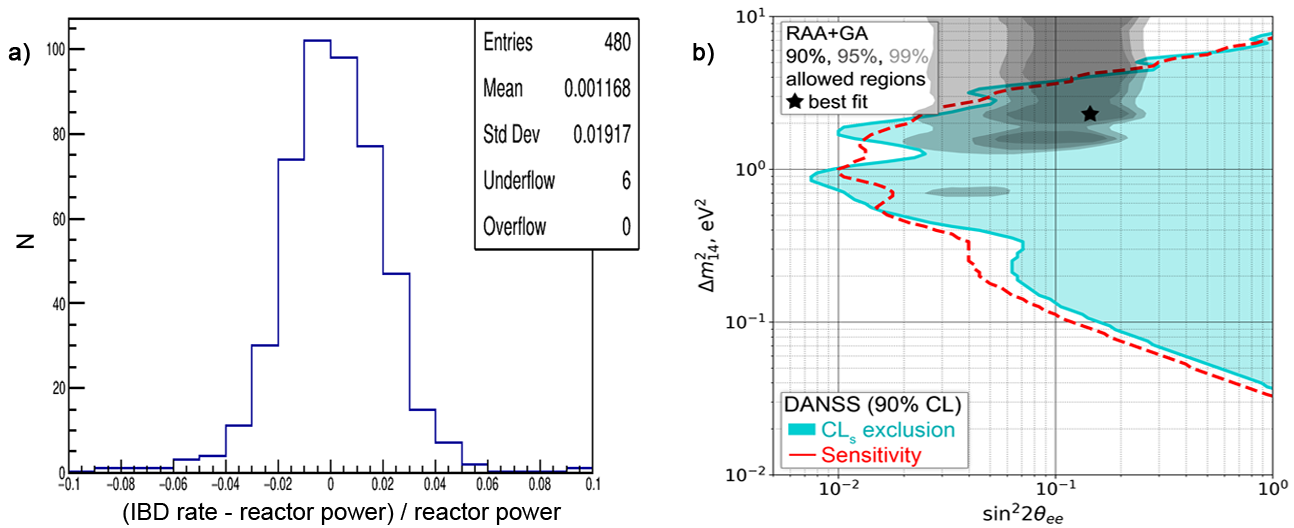}
\caption{a) Relative difference between IBD event rate in about 2 day periods and the reactor power during the reactor operation at full power; b) 90\% C.L. exclusion area (cyan) obtained with the Gaussian CL$_s$ method (preliminary); dashed line is the sensitivity boundary.} 
\label{fig:exclusion}

\end{figure}

For the search of sterile neutrinos only ratios of $e^+$ spectra at the different positions from the reactor core were considered.
Systematic uncertainties in the energy scale and backgrounds were treated as nuisance parameters in the test statistics~\cite{RelEfficiency}. No statistically significant evidence for sterile neutrinos was found. The best-fit point for the 4$\nu$ hypothesis $\mydm = 1.3$eV$^ 2$, $\mysin=0.014$ has the statistical significance of 1.3$\sigma$ only. The exclusion area for the sterile neutrino parameters shown in  Fig.~\ref{fig:exclusion}b was obtained  using a Gaussian \clsmethod\ method~\cite{CLS}. It covers a very large and probably the most interesting fraction of the expectations based on the 
RAA and 
GA~\cite{Mention2011}. 

The collaboration appreciates the permanent assistance of the KNPP administration
and Radiation Safety Department staff.
This work is supported by the Russian Science Foundation grant 17-12-01145$\Pi$.


\end{document}